# VERA_Epidemiology – White Paper 1:
# Using VERA to explain the impact of social distancing on the spread of COVID-19

William Broniec, Sungeun An, Spencer Rugaber & Ashok K. Goel
Design & Intelligence Laboratory, School of Interactive Computing,
Georgia Institute of Technology, Atlanta, Georgia 30307, USA

**Abstract**. COVID-19 continues to spread across the country and around the world. Current strategies for managing the spread of COVID-19 include social distancing. We present VERA, an interactive AI tool, that first enables users to specify conceptual models of the impact of social distancing on the spread of COVID-19. Then, VERA automatically spawns agent-based simulations from the conceptual models, and, given a data set, automatically fills in the values of the simulation parameters from the data. Next, the user can view the simulation results, and, if needed, revise the simulation parameters and run another experimental trial, or build an alternative conceptual model. We describe the use VERA to develop a SIR model for the spread of COVID-19 and its relationship with healthcare capacity.

## 1. Introduction

Newspaper articles in recent weeks have been filled with stories about the spread of COVID-19 across the country and around the world. As the virus continues to spread, various countries are both searching for pharmaceutical mechanisms to combat the disease (such as vaccines and cures) and adopting social strategies for preventing or mitigating its spread (such as social distancing). The mix of mitigation strategies seems to vary among countries depending on factors such as demographics, culture, technology, economic resources, healthcare capacity, and leadership (McCurry, Ratcliffe & Davidson 2020).

On March 14, 2020, Washington Post published an influential article describing how different levels of social distancing may affect the spread of the virus using simple agent-based simulations (Stevens 2020). Figure 1 illustrates the effect. The curve in red shows a rapid increase and decrease in the spread of the virus in a relatively short time period if no precautions are taken. The curve in yellow shows a slower increase and decrease in a relatively longer period if social distancing is practiced. The idea is that a country may adopt a social distancing strategy to "flatten the curve" so that the healthcare capacity of the country is not overwhelmed. Since then, several other studies have established this point more firmly, including a detailed study from Imperial College of London (Ferguson et al. 2020).

In this article, we describe VERA_Epidemiology (or just VERA for short), an interactive AI tool that enables users to build conceptual models of the impact of social distancing on COVID-19. Unlike some of the other simulations, VERA enables the user to explicitly specify the conceptual model in a visual language and automatically spawns agent-based simulations from the conceptual models. Given a set of data, VERA automatically extracts initial values for the simulation parameters from the dataset, and it also affords the user to interactively revise the parameter values. The user can view the simulation results, and, if needed, revise the simulation parameters and run another experimental trial, or build an alternative conceptual model. We demonstrate the model development process through comparative models of the impact of social distancing using the Johns Hopkins University dataset (CSSEGISandData, 2020).

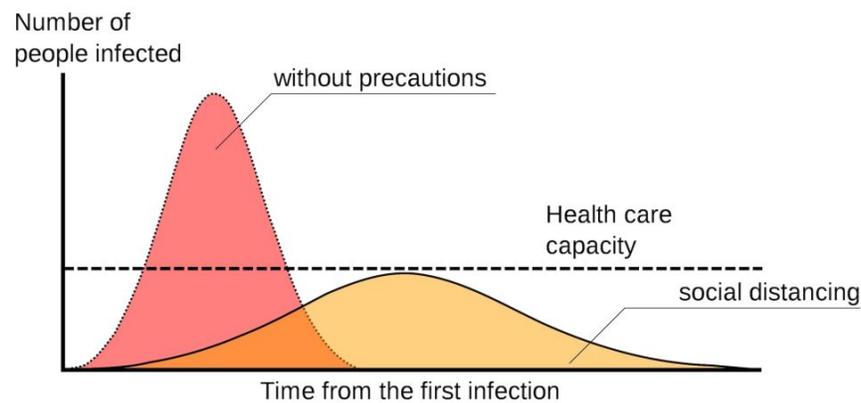

*Figure 1:* *The effect of social distancing on the spread of COVID-19. Flattening the curve helps keep the number of people infected with the virus within the range of the healthcare capacity. (Adapted from Stevens 2020, published in The Washington Post on March 14, 2020.)*

We describe the use of VERA to develop a SIR model for the spread of COVID-19 and its relationship with healthcare capacity. The results show gradual "flattening of the curve" as increasingly intense social distancing strategies are implemented. More importantly, VERA acts like a virtual laboratory to conduct "what if" experiments with different SIR models without requiring any knowledge of mathematical equations or computer programming.

## 2. An Illustrative Example of Modeling the Spread of COVID-19 Using VERA

The Virtual Experimentation Research Assistant (VERA) is a web application that enables users to construct conceptual models of complex systems and run model simulations. The original VERA system operated in the domain of ecological systems and has been extensively used for learning and education in ecology (An et al. 2018; http://vera.cc.gatech.edu). The VERA system described here is an adaptation of the original VERA for ecological modeling.

We illustrate the use of VERA to create three models that form a series of increasingly intense social distancing policies (see Figure 2). We use two parameters to control the levels of social distancing: *interaction probability* and *adoption/transmission interval.* These two parameters are incrementally increased from Model 1 to Model 3 to increase the interaction probability while the COVID-19 spread component itself remains intact.

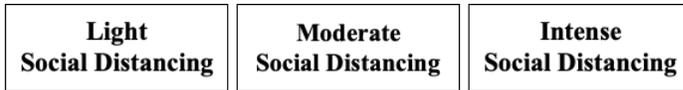

*Figure 2. Three models with increasing levels of social distancing.*

## 2.1 Conceptual Modeling

VERA uses conceptual models to enable users to visually express the components and relationships of a system. Figure 3 illustrates a screenshot from VERA's editor for constructing conceptual models in epidemiology. The left panel contains a palette for adding different types of components to the model. The grid panel in the center is where the conceptual model is assembled. The right panel depicts model parameters and their initial values for the simulation of the conceptual model. Note the visual nature of the conceptual models in VERA. The conceptual models in Figure 2 illustrate an interaction between social distancing and COVID-19 cases.

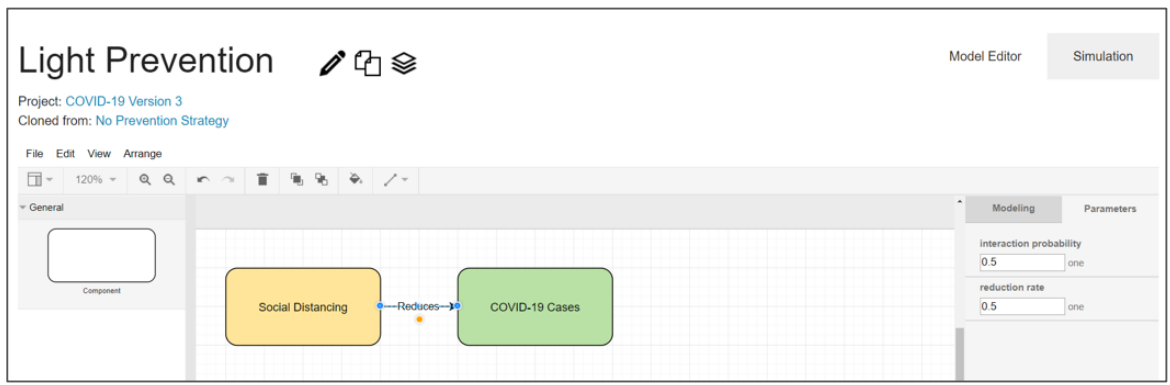

*Figure 3:* Conceptual modeling in VERA

## 2.2 Simulation Parameters

VERA translates qualitative conceptual models into quantitative values and equations to generate agent-based simulations. VERA currently uses the following parameters specific to epidemiology: *starting count, duration, adoption/transmission count, adoption/transmission onset, and adoption/transmission interval.* VERA facilitates describing the simulation parameter values in two ways. First, VERA enables users to import data from external sources and then uses machine learning techniques for abstracting initial simulation parameter values from the data. Second, it enables users to interactively set the parameter values.

Table 1 shows increasing the interaction probability and the adoption /transmission interval parameters to represent the increasing level of social distancing. The values of the simulation parameters of the *COVID-19 Cases* component in Figure 4 were automatically filled in by applying machine learning techniques on the imported dataset (CSSEGISandData, 2020).

*Table 1: Increasing the interaction probability and the adoption /transmission interval parameters to represent increasing level of social distancing.*

|  | **Light Social Distancing** | **Moderate Social Distancing** | **Intense Social Distancing** |
|---|---|---|---|
| *Interaction Probability* | 0.5 | 0.71 | 0.84 |
| *Interval* | 12 | 25 | 28 |

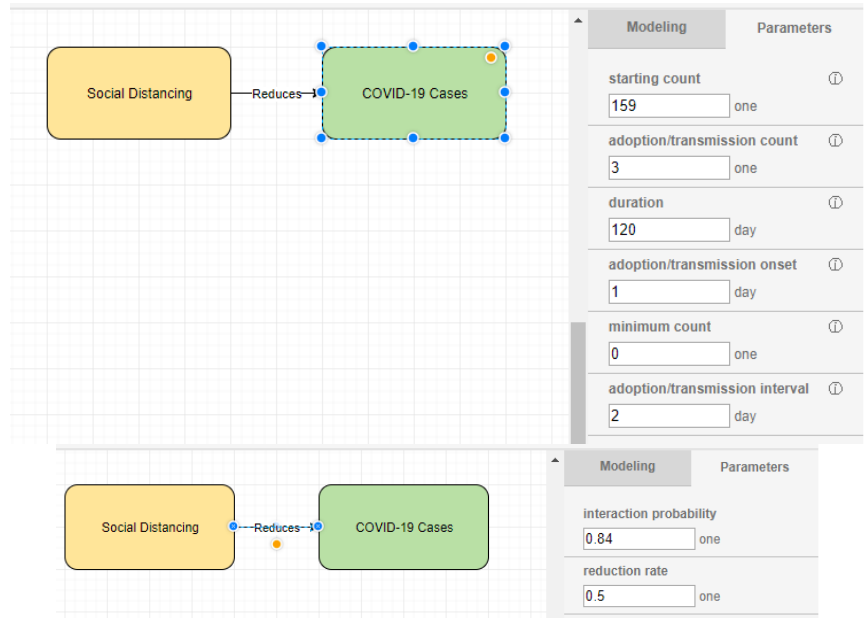

*Figure 4: The simulation parameters of the COVID-19 CASES components.*

## 2.3 Simulation Results

VERA uses an off-the-shelf agent-based simulation system called NetLogo (http://ccl.northwestern.edu/netlogo/). Running the simulation enables the user to observe the evolution of the system variables over time and iterate through the generate-evaluate-revise loop.

Figure 5 shows the simulation results generated from the model in Figure 4 using different parameter values, as indicated in Table 1. The first model shows one future progression where the rate spikes early, but there is a sharp decline. This can be compared to South Korea's handling of COVID-19 spread, where medical technology and mass screenings drastically limited the spread. The primary concern here is the capacity of the healthcare system, where demand can easily exceed supply. The next graph shows a model of moderate social distancing. In this case, the peak occurs later, does not reach as high, and stretches the pandemic out for a longer period. Such approaches are straining the healthcare capacity of many European nations now and may not be enough. The final model shows the results of intense social distancing on infection rates. These strong lifestyle changes drastically reduce the number of potential avenues for disease spread, and the resulting spread of infections occurs much slower, providing time, resources, and planning for a society to operate in the times of a pandemic.

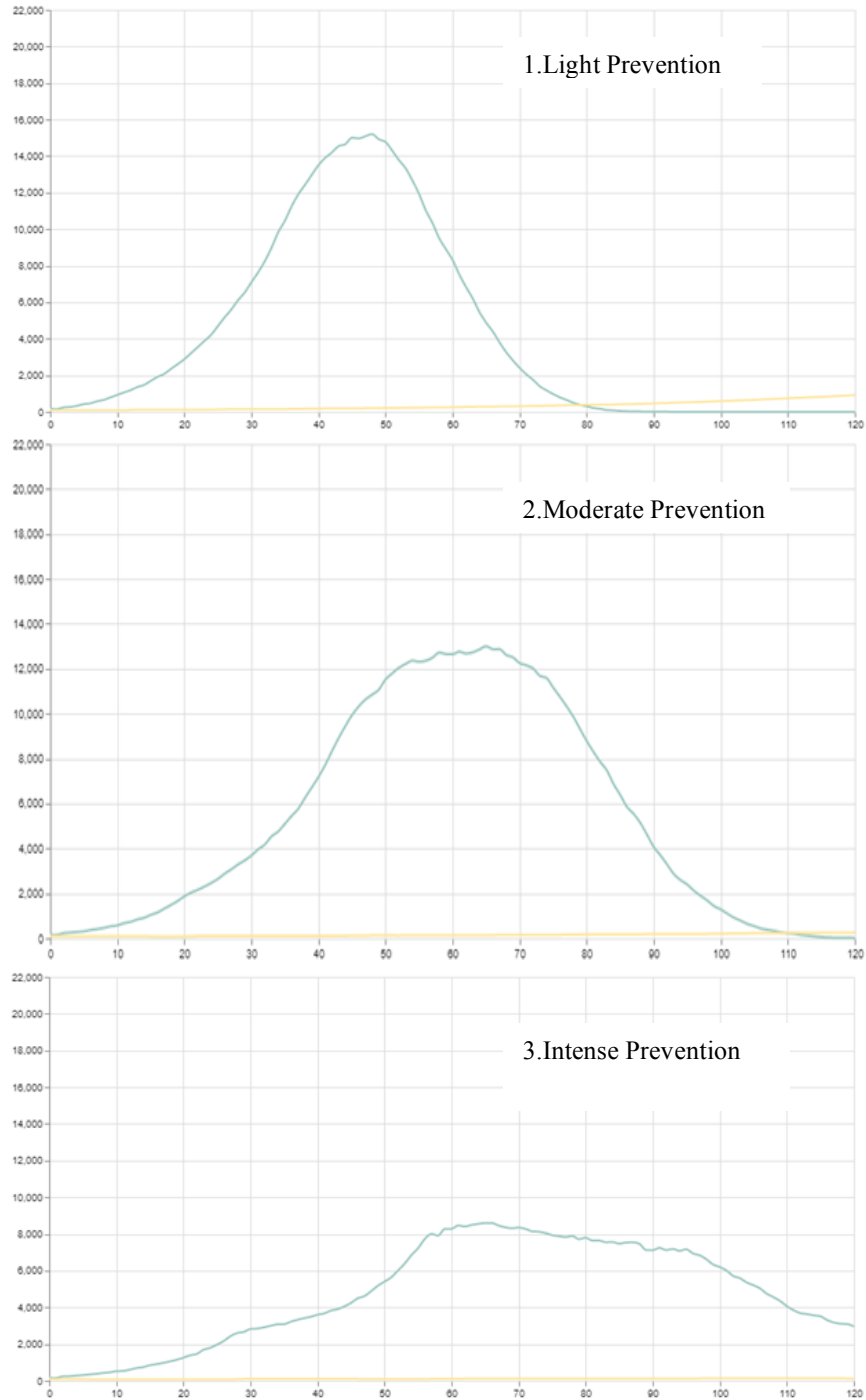

*Figure 5:* *The simulation results of Models 1-3 (top to bottom).*

## 3. The SIR Model of COVID-19 Spread in VERA

Now that we have illustrated the core techniques in VERA, we describe the use of VERA develop the SIR model for understanding the relationship between social distancing and the spread of COVID-19. The SIR model (Kermak & McKendrick 1927) is a commonly used mathematical model to understand the spread of infectious diseases. We show just how important adjusting behavior can be in shaping the outcome of a pandemic. The strain on the healthcare infrastructure has sweeping consequences—reduced availability of resources prevents some people from receiving adequate treatment, and even those with conditions other than COVID-19 will still see negative impacts as these resources become scarcer.

### 3.1 Conceptual Modeling

The conceptual model in Figure 5 illustrates the SIR model of disease spread. The conceptual model consists of three components that each represent a number of individuals—the susceptible, the infected, and the recovered—as well as the healthcare capacity. The susceptible population becomes the infected population, and the infected population becomes the recovered population.

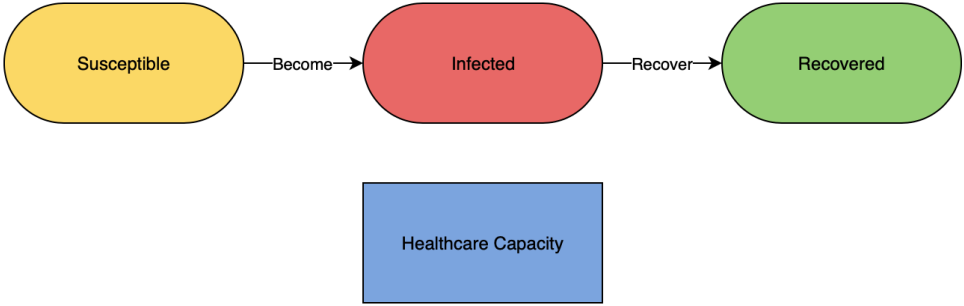

*Figure 5:* *The SIR conceptual model in VERA*

### 3.2 Simulation Parameters

For the SIR model, we use a different set of parameters than for the simple model described in the previous section. The *susceptible*, *infected* and *recovered* components have *starting populations* as the simulation parameter. The healthcare capacity component has *capacity* as the simulation parameter. The *Become* relationship has two simulation parameters: *Average contacts per day per person* and *Transmission Likelihood*. The *Recover* relationship has *Average recovery time* parameter (see Table 2 for description).

*Table 2*: *The Simulation Parameters of Become and Recover Relationships.*

| Relationship | Simulation Parameter | Description |
|---|---|---|
| Become | *Average contacts per day per person* | Describes how many people the average person is coming in contact with on a given day in this simulation. |
|  | *Transmission Likelihood* | Describes the probability of the disease transferring from an infected person to a susceptible person that they have come in contact with. |
| Recover | *Average recovery time* | Describes how long it takes in average to recover from the disease. |

### 3.3 Simulation Results

Figure 6 shows the simulation results on a sample population of 10,000 people with two different *Average contacts per day per person*. The first model (top) shows the simulation results with 16 *Average contacts per day per person*. The red series corresponds to the number of infected individuals, and it exceeds the healthcare capacity of the system very early under these conditions. The second model (bottom) shows the simulation results with 12 *Average contacts per day per person*. Reducing the *Average contacts per day per person* hypothetically suggest that people are reducing social contact somewhat, but not substantially. Compared to the prior graph, the peak is closer to 7,000 than 8,000, and healthcare capacity is exceeded after 20 days in the simulation, rather than around 15 days—a step in a positive direction. The users can experiment with other values of *Average contacts per day per person* until they find a scenario where infections do not exceed healthcare capacity.

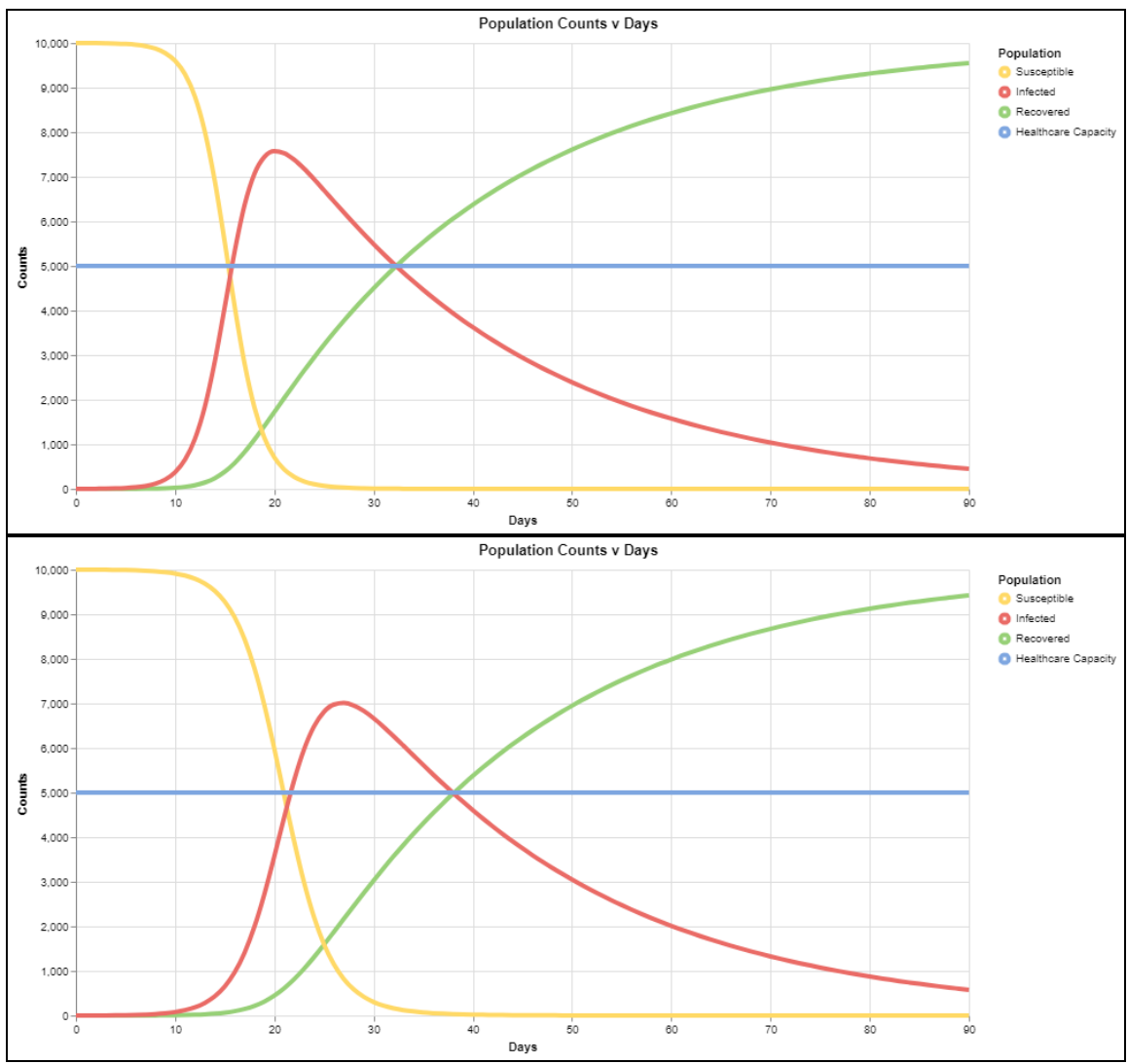

***Figure 6****: Simulation Results with 16 Average contacts per day per person (top) and 12 Average contacts per day per person (bottom).*

## 4. Reflections

In this article, we used VERA to develop SIR models for the spread of COVID-19. These models not only express the impact of social distancing on the spread of the disease but also the management of the impact of the disease on the healthcare capacity. We note, however, that VERA supports conceptual modeling, not mathematical modeling, and that its models are more explanatory of general patterns than predictive of specific outcomes.

VERA (1) enables the user to explicitly specify the conceptual model in a visual language, (2) automatically spawns agent-based simulations from the conceptual models, (3) automatically extracts initial values for the simulation parameters from a given dataset, and (4) supports the user through the whole cycle of model generation, evaluation and revision. Thus, VERA provides a virtual laboratory: the user can try out a variety of conceptual models and simulation parameters, and conduct "what if" virtual experiments. We posit that this has significant implications for learning and education, for example, informal learning by the citizens of the world.

**Acknowledgements:** We thank Preethi Sethumadhavan for her contributions to this work. The development of the original version of VERA for ecological modeling was supported by an NSF BigData grant. We thank Robert Bates (Veloxicity), Dr. Jennifer Hammock (Encyclopedia of Life, Smithsonian Institution), and Dr. Emily Weigel and Brady Young (School of Biological Sciences, Georgia Tech) for their contributions to the original VERA project.